	\documentclass[namedreferences]{solarphysics}
\usepackage[hyperref,optionalrh,solaromanenum]{spr-sola-addons} 
\usepackage{graphicx}                    
\usepackage{color}                       
\usepackage{breakurl}                         

\newcommand{\pref}{\protect\ref}

\begin{document}

\begin{article}

\begin{opening}

\title{The Standardisation and Sequencing of Solar Eclipse Images for the Eclipse Megamovie Project}

%
\author{Larisza D.~Krista{}$^{1, 2, 3}$\sep
        Scott W.~McIntosh{}$^{2}$\sep
       }

%
\runningauthor{L. Krista \& S.W. McIntosh}
\runningtitle{Standardisation of Solar Eclipse Images}

%
  \institute{$^{1}$ Cooperative Institute for Research \\in Environmental Sciences,\\University of Colorado at Boulder,\\ Boulder, CO 80309, USA\\
                     email: \url{larisza.krista@colorado.edu}\\ 
             $^{2}$ High Altitude Observatory,\\ National Center for Atmospheric Research,\\ P.O. Box 3000,\\ Boulder, CO 80307, USA\\
                     email: \url{mscott@ucar.edu} \\
             $^{3}$Space Weather Prediction Center,\\ National Oceanic and Atmospheric Administration,\\ Boulder, CO 80305, USA}

\begin{abstract}
We present a new tool, the {\it Solar Eclipse Image Standardisation and Sequencing} (SEISS), developed to process multi-source total solar eclipse images by adjusting them to the same standard of size, resolution, and orientation. Furthermore, by analysing the eclipse images we can determine the relative time between the observations and order them to create a movie of the observed total solar eclipse sequence. We successfully processed images taken at the 14 November 2012 total solar eclipse that occurred in Queensland, Australia, and created a short eclipse proto-movie. The SEISS tool was developed for the Eclipse Megamovie Project (EMP: \href{https://www.eclipsemegamovie.org}{www.eclipsemegamovie.org}), with the goal of processing thousands of images taken by the public during solar eclipse events. EMP is a collaboration among multiple institutes aiming to engage and advance the public interest in solar eclipses and the science of the Sun--Earth connection.
\end{abstract}

%

\end{opening}

%
 \section{Introduction}
 \label{intro} 
Total solar eclipses have captured the imagination of the human race for millennia. While they were immortalized in oral histories and on dwelling walls, we now see them as a fleeting glance at the Sun's enigmatic corona. Solar eclipses also motivated the design of an entire class of solar instrumentation to provide eclipse-like conditions all the time \citep[{\it e.g.} ][]{Lyot39}. Contemporary total solar eclipse experiments \citep[{\it e.g.} ][]{Pasachoff15} provide routine opportunities to explore new wavelength domains and camera technologies. The Eclipse Megamovie Project (EMP) has a complementary purpose -- it aims to engage the general public in the day--to--day connections between the Sun and Earth. Citizen scientists can then use their cameras to image the progression of the total solar eclipse from their location, upload their images through their device, or through a webpage, and contribute to a movie of the country's collective observation of the eclipse progression. The upcoming total solar eclipses that cross the continental US (in 2017 and 2024) motivate the EMP project and the development of the image pipeline discussed herein.

The primary challenge of EMP is to take a heterogeneous set of eclipse images, co-register them, and sequence them into a seamless movie of the solar eclipse evolution. We have developed the {\it Solar Eclipse Image Standardisation and Sequencing} (SEISS) algorithm, which uses advanced image processing techniques to select high-quality solar eclipse images and prepare them for the eclipse-movie sequence. The observations in our test image database were made during the 14 November 2012 total solar eclipse that occurred in Queensland, Australia. The images came from a wide variety of observers and cameras, and our database contained eclipse observations of different resolution and orientation, as well as random holiday photos. Our goal was to process this database in the most efficient and automated fashion.

SEISS was developed to be a quick, efficient, robust algorithm that can process thousands of submitted eclipse photos. Furthermore, SEISS is the foundation for a future tool that will operate on individual devices (such as smart-phones), and in a server-mode, where there will be instant feedback to the user by sequencing and packaging {\em their} images and constructing a movie. DSLR cameras and smart-phones are less likely to produce research-level quality imaging sequences, hence, we minimised computationally heavy and high-level processing to maintain a robust and fast tool. Nevertheless, for eclipse observations carried out by professional astronomy groups ({\it e.g.} the Citizen Continental America Telescopic Eclipse (CATE) Experiment) using stably mounted larger aperture image-capture devices, we will further enhance SEISS capabilities, and explore other techniques to capitalise on existing eclipse image-analysis techniques \citep[{\it e.g.} ][]{Druckmuller09, Morgan14}. 

We developed the SEISS tool in preparation for the 21 August 2017 eclipse which will be a major event with considerable public interest. The path of the eclipse totality will traverse the entire breadth of the continental United States, from Oregon to South Carolina. It will be the first of its kind since 1918. It  will provide scientists with a unique opportunity to assemble a very large number of eclipse images, obtained by observers all along the path of totality, into a continuous record of coronal evolution during the event. The coast--to--coast transit will take about 95 minutes -- allowing us to build an ``Eclipse Megamovie'' \citep{Hudson11, Davey13}. This will be a multi-observer, ``citizen science'' movie of the total solar eclipse that can capture the morphology and evolution of the solar chromosphere and corona. This event, and a similarly spectacular event on 8 April 2024, will provide an important opportunity for public education and outreach about the Sun and its interaction with the Earth.
 
The following sections of this article lay out a prototype image pipeline for the SEISS algorithm.
 
\section{Method}
In this section we discuss the image processing steps used to identify the solar disk in the eclipse images, to standardise the images ({\it e.g.} size, alignment) and to order them temporally with the ultimate goal of creating a sequenced movie of the entire total solar eclipse.

\subsection{Solar Disk Detection Using the Hough Transform}
\label{circID}
In order to detect circles in an eclipse image, we first enhance the edge of the solar disk. This is done by applying a Sobel filter to the image -- a two-dimensional spatial gradient measurement that permits identification of edges in an image.  (For more information on the Sobel operator see the following link: \url{https://www.researchgate.net/publication/210198558\_A\_3x3\_isotropic\_gradient\_operator\_for\_image\_processing})

All edges in the eclipse image are enhanced by this step -- not only the edge of the solar disk. To eliminate false disk detections we employ a modified Hough transform to identify the centres of all circles present in the image, including that of the solar disk. The properties of the edges can then be used to identify that belonging to the Sun, which has the most ``prominent edge''. The discrimination process starts with the creation of an accumulator space (or Hough space) where each cell in the space corresponds to a pixel in the image. Next, a circle is fitted to each edge pixel. The more circles overlap at a pixel location the higher accumulation it obtains (i.e. it becomes a higher-value accumulation cell). The pixel that corresponds to the highest-value accumulation cell defines the centre of the circle with the largest number of edge points -- the most prominent edge and circle. The less prominent circles can also be identified; these will correspond to the lower-value accumulation cells.

\begin{figure}[!t]
\centerline{
\includegraphics[scale=0.45, trim= 0 70 0 50 , clip]{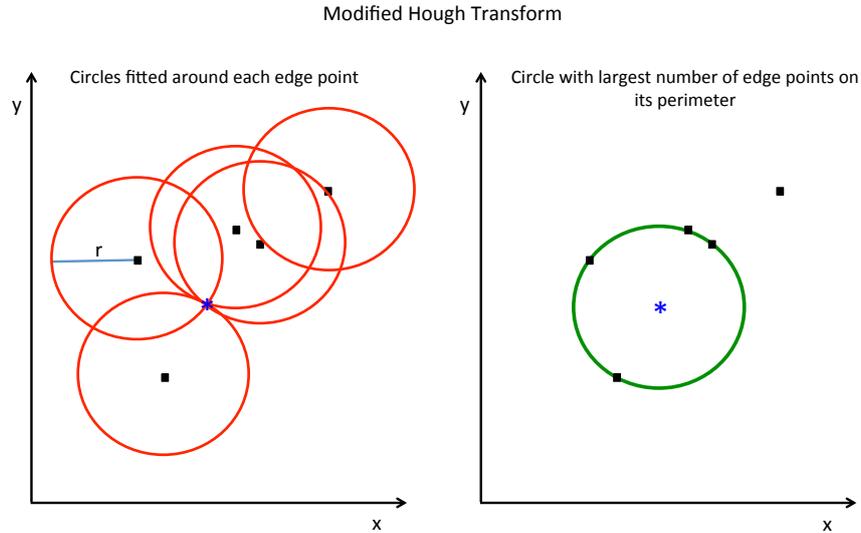}}
\caption{Finding the centre of the circle (of radius $r$) with the largest number of edge pixels on its perimeter. Left: the five black points denote the most prominent edge pixels. A (red) circle of radius $r$ is drawn around each edge pixel. The location with the highest number of intersecting circles is highlighted by a blue star. This location is the centre of the most prominent circle. Right: a (green) circle or radius $r$ is drawn around the (blue star) center location, showing that four edge pixels fall on the most prominent circle.} 
\label{houghcirc}
\end{figure}

Let us demonstrate the process with a simple example: we look for a single circle of radius $r$ in an image that contains five edge points (Figure~\ref{houghcirc}). The circle to be identified is one that contains the largest number of edge points on its perimeter and has a radius $r$. In the left panel of Figure~\ref{houghcirc} (red) circles of radius $r$ are drawn around each edge pixel (black points). The overlapping points between the red-circle perimeters define the centres of possible circles. The pixel containing the highest number of perimeter overlaps ({\it e.g.} the highest frequency point in the accumulator space) is then defined as the centre of the most prominent circle: the circle with the largest number of edge pixels on its perimeter.

In our case the highest-value accumulation cell (pixel) is shown with a blue star. We fit a circle of radius $r$ to this point to highlight the most prominent circle that has four edge pixels on its perimeter (shown as a green circle on the right side of Figure~\ref{houghcirc}).

\begin{figure}[!t]
\centerline{
\includegraphics[scale=0.6, trim= 50 120 50 150 , clip]{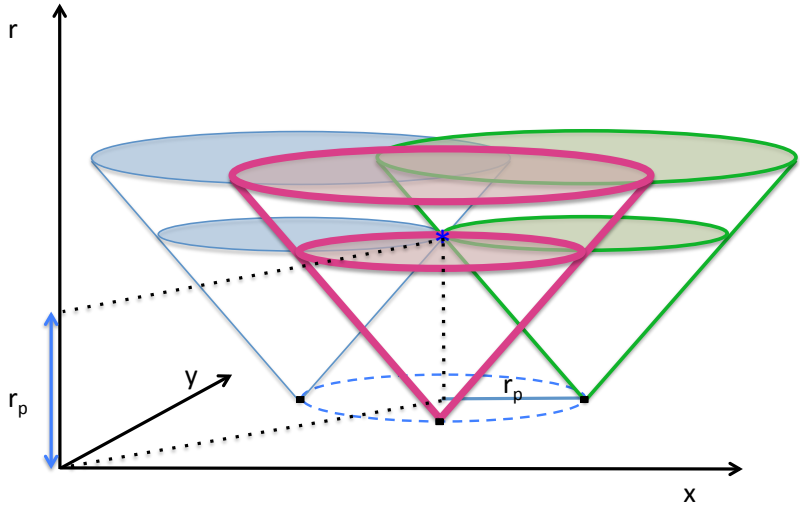}}
\caption{Identifying the most prominent circle radius using cones. The $x$--$y$-plane is the image plane, and the third dimension is radius. Three edge pixels are shown on the $x$--$y$-plane (black points). Each of these pixels is fitted with a cone (red, green and blue). The point where the cones intersect (blue star) identifies the radius [$r_{p}$] of the most prominent circle (blue dashed circle) that fits all three points on its perimeter.}
\label{cones}
\end{figure}

The detection of the most-prominent circle using this method is relatively fast and easy when the radius of the circle is known, as the accumulator space is two-dimensional. However, for a heterogeneous eclipse image set, we have no information on the solar-disk radius in each image; we must identify it. For our generalised purpose we have to use a three-dimensional accumulator space, where the third dimension allows the use of a range of possible radii (see Figure~\ref{cones}). In this case each edge pixel in the image is fitted with a (three-dimensional) cone rather than a (two-dimensional) circle. The centre and radius of the most-prominent circle in the image is determined by the point (or accumulation cell) where the largest number of cone surfaces intersect (highlighted with a blue star in the figure). To keep our example simple, in Figure~\ref{cones} we only show three edge pixels (black points on the $x$--$y$-plane). Each of these pixels is fitted with a cone. The third dimension in this coordinate system corresponds to the radius -- the point where the cones intersect (blue star) identifies the radius [$r_{p}$] of the circle that has all three edge points on its perimeter (shown as the blue dashed circle on the $x$--$y$-plane).

In our study we used the IDL algorithm \textsf{circlehoughlink.pro} to identify the most prominent circles. This algorithm is available at the following link: 
\newline
\url{https://www.cis.rit.edu/class/simg782.old/talkHough/circleHoughPrograms.html}.

\subsection{Image Standardisation}

\begin{figure}[!t]
\centerline{
\includegraphics[scale=1.25, trim= 30 260 50 110 , clip]{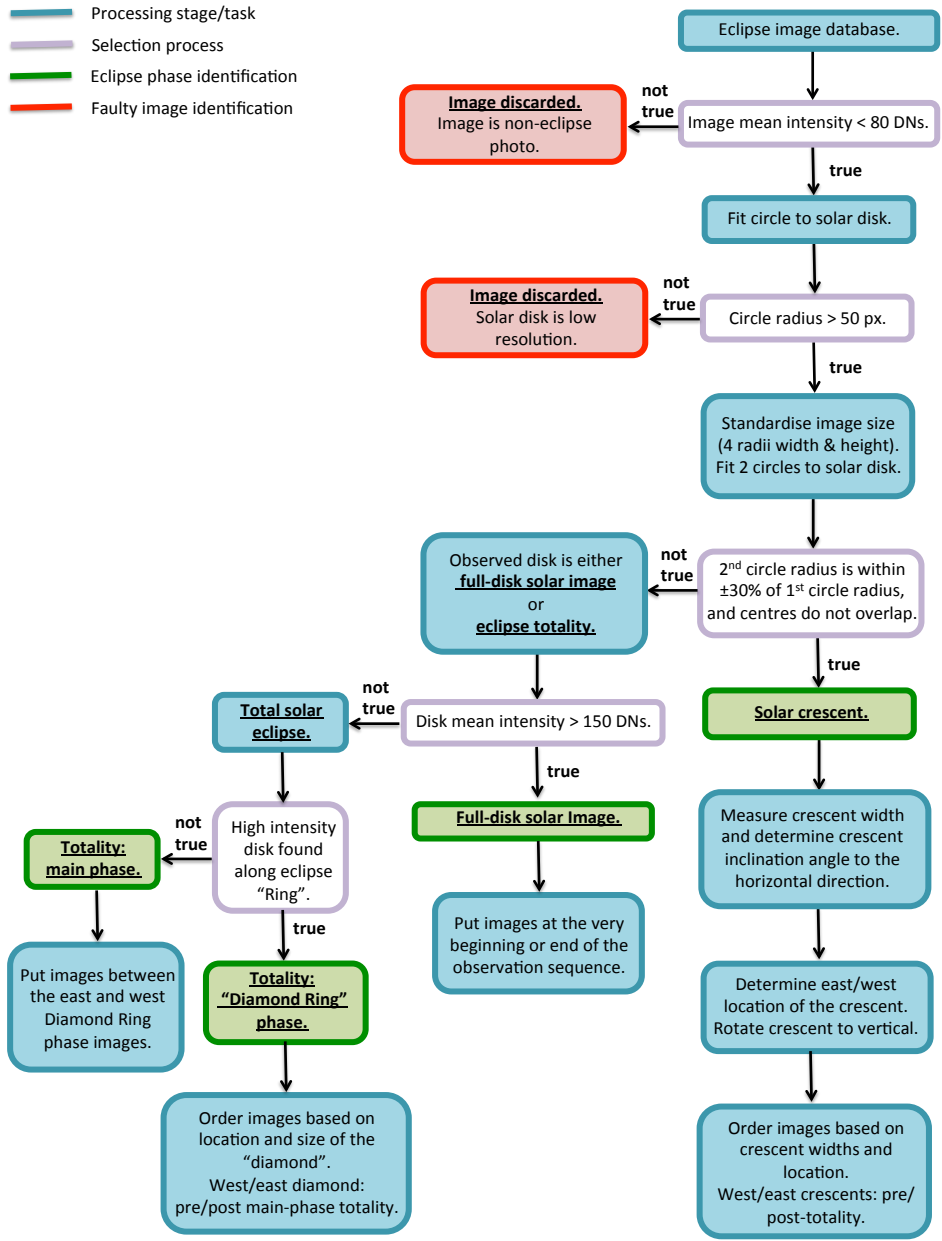}}
\caption{The image processing steps used in the SEISS tool.}
\label{flowchart}
\end{figure}

Following the identification of the solar disk in each eclipse image, the next step is to standardise the images by converting them to the same size, resolution and orientation. This is followed by arranging the images in the correct temporal order to create a seamless solar eclipse movie. Our goal was to create a robust method that can filter and process a large variety and quality of images made by different instruments. Our test database contained 282 images in total (from approximately 20 different cameras). Many of these images were non-eclipse images. After filtering out non-eclipse images and low-quality eclipse images, we had 63 images from which to make the proto-movie. A flowchart of the processing steps is shown in Figure~\ref{flowchart}. 
 
\subsubsection{Filtering Out Unwanted Images}
We start the SEISS process by filtering out unwanted images. We categorised these as:
\begin{itemize}
\item{{\bf Irrelevant images}: {\it e.g.} photos of pretty beaches, people, cats, etc.
These images were filtered out using an intensity threshold: if the mean intensity of the image was higher than 80DNs (Data Numbers -- ranging from 0 -- 255 in the byte-scaled images), the image was discarded. The threshold was set based on eclipse and non-eclipse image intensities in our database -- non-eclipse images typically have higher mean intensity than eclipse images.}
\item{{\bf Faulty or low-quality observations}: {\it e.g.} an image where a cloud partly obscured the eclipse crescent or the eclipse crescent was low resolution. If the intensity of the image was low enough to pass through the first (intensity) filtering step, a circle-fitting filter detected the faulty images: we fitted a circle in the image (as described above) and if the radius was too small ($\leq50$ pixels) the image was discarded. The circle-fitting tool attempts to fit prominent edges -- ideally those belonging to a solar disk or crescent. Typically, in low-resolution images, the detected Sun has a small radius and hence the image is discarded. If there is no large circular object (Sun) in the image, the prominent edges are other small features which would be fitted with small circles (and consequently the image would be discarded). It is possible that there could be a large enough round feature in an image which could be mistaken for the Sun: {\it e.g.} a beach ball. How would our method discriminate such an image? A beach ball would typically be photographed in a well-lit environment, so the intensity filter would automatically discard it. Hence, unless the false object is a luminescent beach ball in a dark environment it will be detected by our filters and excluded from further steps.}
\end{itemize}

\subsubsection{Differentiating Between Solar Crescents, Eclipse Totality, and Full-Disk Observations}

The intensity and radius filters were found adequate to filter out unwanted images. The high-quality images were then analysed to determine if the image was a solar crescent, totality,  or a full-disk observation. This disambiguation step was performed by fitting two circles in the studied image. In the case of a solar crescent the inner and the outer edge of the crescent could be fitted with two separate circles. In the case of eclipse totality and full-disk solar images, a suitable second circle was not found. The totality and the full-disk images were then distinguished by the mean intensity of the pixels inside the fitted circle: a full-disk image was required to have a mean intensity above 150~DNs, otherwise the observation was classed as a total solar eclipse image.

\begin{figure}[!t]
\centerline{
\includegraphics[scale=0.55, trim= 0 270 0 120, clip]{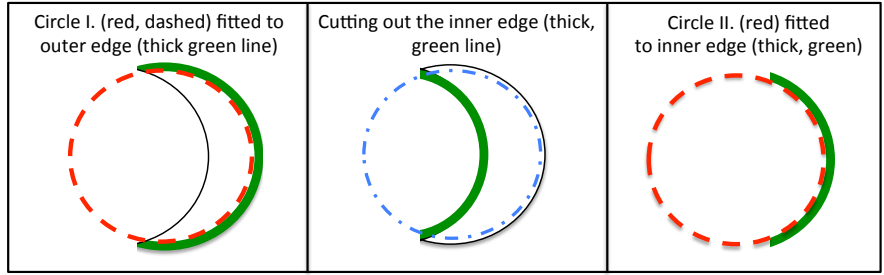}}
\caption{Fitting circles to the outer and inner edge of the solar crescent. The solar crescent is shown with a thin, solid black line. Left: The first circle (red-dashed line) is fitted to the outer edge (thick, solid green line) of the crescent. Middle: The inner edge (thick, solid green line) of the crescent is separated from the outer edge by cutting out a disk (blue, dot-dashed line) that has a 7\,\% smaller radius than the original. Right: The inner edge (thick, solid green line) is fitted with the second circle (red dashed line).}
\label{double_circ}
\end{figure}

The solar eclipse crescents were fitted with two circles. The algorithm discussed above (\textsf{circlehoughlink.pro}: Section~\pref{circID}) primarily detects the circular object with the largest number of edge pixels. Since the outer edge of the crescent is always longer, it is always the dominant circle detected by the algorithm (and the next best circles are also various similar fits to the outer edge). To tackle this problem, we extracted the inner-crescent edge, by cutting out a disk with a 7\,\% smaller radius than the original-circle radius, as illustrated in Figure~\ref{double_circ}. The outside of that circle was then masked and the inner edge of the crescent was fitted with a circle. Finally, once the outer and inner edges of the crescent were successfully identified and fitted, the distance between the two circle centres was measured to determine the thickness of the crescent; this thickness was later used for the temporal ordering of the crescent images.

\subsubsection{Rotation of the Eclipse Observations to a Standard Alignment}
\label{rotate}

\begin{figure}[!t]
\centerline{
\includegraphics[scale=0.4]{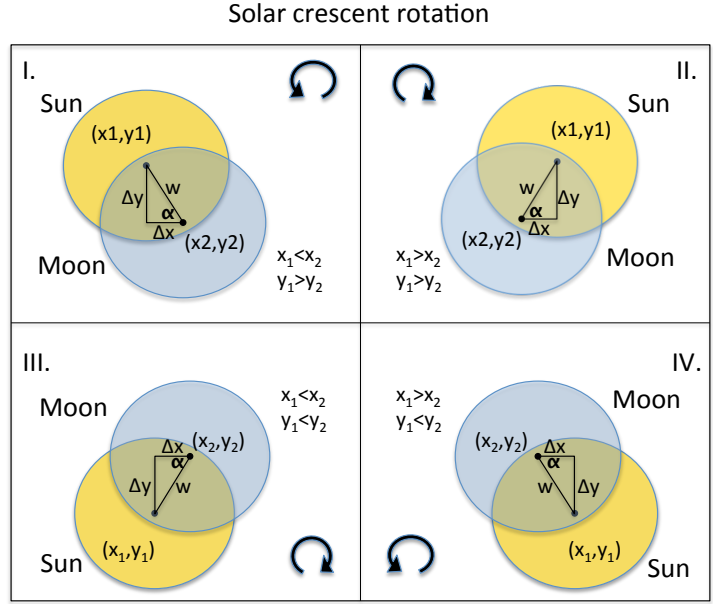}}
\caption{The rotation of the solar crescent for the standardised alignment. The rotation direction is determined based on the relative location of the Sun and the Moon: in Case I and IV the crescent is rotated anti-clockwise, in Case II and III the rotation is clockwise. The rotation angle is $\alpha$, and the crescents are rotated so the peaks line up with the N--S direction.}
\label{solar_rot}
\end{figure}

The line connecting the centres of the two crescent circles was used to determine the inclination angle [$\alpha$] relative to the horizontal (see Figure~\pref{solar_rot}). The coordinates $x_{1}$ and $y_{1}$ denote the centre of the circle fitted to the Sun, and $x_{2}$ and $y_{2}$ define the centre of the circle fitted to the Moon. The distance between the centroids is given by $w=\sqrt {{\Delta x}^2{\Delta y}^2 }$, where $\Delta x=x_{2}-x_{1}$ and $\Delta y=y_{2}-y_{1}$. The distance $w$ also equals the width of the crescent. 

The rotation angle is then given by $\alpha$=asin${\frac{\Delta y}{w}} $ and is used to de-rotate the images so that all of the crescents are oriented in the same fashion, aligned with the North--South direction. The rotation direction (clockwise or anti-clockwise) is dependent on the relative location of the two circle centres. The crescents that are originally in the upper-right or lower-left quadrant of the Sun are rotated clockwise (positive rotation), while crescents in the upper-left and lower-right quadrant are rotated anti-clockwise (negative rotation). Whether the rotation is positive or negative is determined by the sign of the product of $\Delta x \Delta y$. In Figure \ref{solar_rot} we show the four different scenarios of rotation. In all four panels the Sun is shown as a yellow disk, while the Moon is a blue disk. The disks overlap to create the solar eclipse crescent (bright yellow crescent). The positive or negative rotations are shown with a black arrow for each scenario, depending on which quadrant the crescent is located in. {\it E.g.} in the upper left panel (Scenario I) the crescent is located in the upper-left quadrant. As shown, $x_{1} < x_{2}$ and $y_{1} >y_{2}$, which means that $\Delta x \Delta y$ is a negative number, hence the crescent has to be rotated by $\alpha$ in an anti-clockwise direction. $\Delta x \Delta y$ is also negative in Scenario IV, leading to anti-clockwise rotation. In Scenario II and III  the product of $\Delta x \Delta y$ is a positive number and hence the rotation is clockwise.

\begin{figure}[!t]
\centerline{
\includegraphics[scale=0.7, trim= 0 230 0 210, clip]{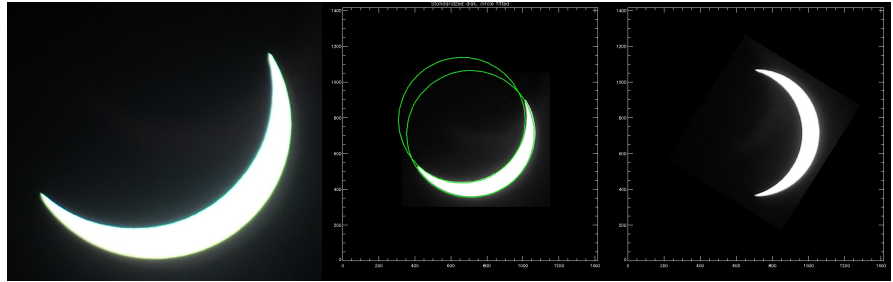}}
\caption{The processing steps used for solar-crescent observations. Left: the original solar crescent observation (this example is pre-totality). Middle: the solar and lunar disks are identified on the standard size image (green circles). Right: the finalised image after rotation. }
\label{crescent}
\end{figure}

In Figure~\ref{crescent} we show the above discussed processing steps on an actual observation. In the left panel, the original, raw observation is shown. This image was resized to the standard format, and then the solar and lunar disks were identified (green circles, middle panel). Finally the image was rotated to the standard alignment using the above described methods (right panel).

\subsubsection{Identifying the Diamond Ring Phase of the Eclipse}

\begin{figure}[!t]
\centerline{
\includegraphics[scale=0.5, trim= 0 130 0 130 , clip]{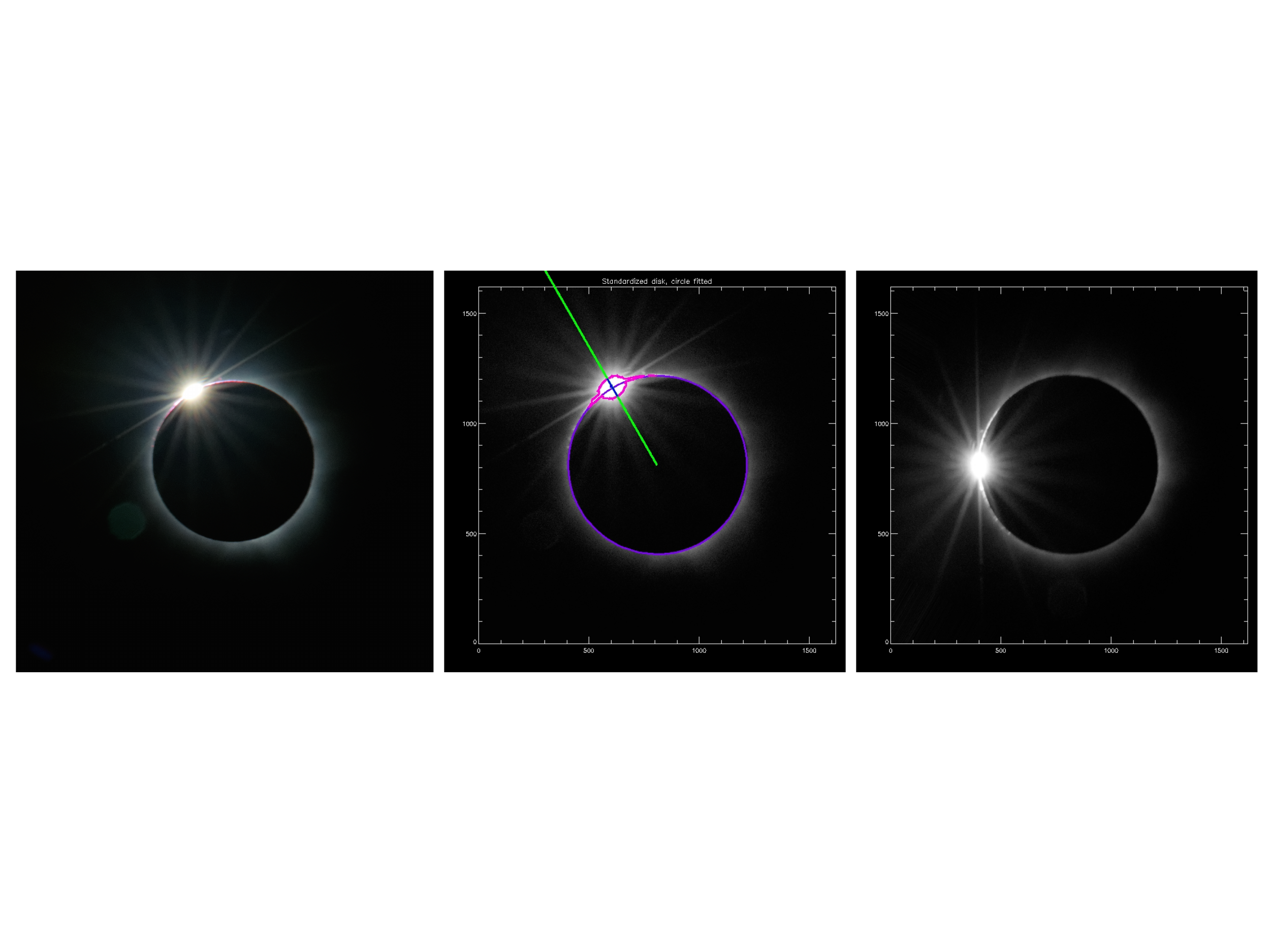}}
\caption{The Diamond Ring phase of the total solar eclipse -- the first/last glimpse of the Sun after/before totality. Left: a raw photo of the diamond ring phase (this example is post-totality). Middle: the analysis of the image. The purple contour shows the circle fitted to the solar disk (the ring). The pink contour highlights the identified diamond feature. The green line connects the disk centre to the diamond centroid and allows the determination of the inclination angle and the diamond width. Right: the final, standardised image. }
\label{diamond}
\end{figure}

We were also able to identify and isolate the images that contain the ``Diamond Ring'' (DR) phase of the total solar eclipse: the first and last moments of the totality ({\it e.g.} see Figure~\ref{diamond}). The ``diamond'' is caused by the last fragment of photospheric light shining from behind the Moon, while the ``ring'' is the solar atmosphere that becomes visible as the sky darkens during the eclipse. The latter remains visible after the DR-phase is over, and the totality is reached. As discussed above we differentiated between images containing totality or crescent information by fitting a single circle or two (non-concentric) circles, respectively. The next step was to distinguish between totality and the DR-phase images. These two phases are similar except the DR-phase images contain the diamond-feature along the ring which is so bright that it saturates onto the disk. 

The requirement for identifying the DR-phase images was a low-intensity main disk with a very bright feature along the circle ($>$200 DN) that ``spilled over'' onto the disk. Note that totality images also have bright features along the circle (features in the solar corona) but those do not extend onto the disk. Once the diamond was identified, it was contoured (by intensity) and its centroid was determined via the centre of mass method. Next, a line was drawn from the centre of the solar disk through the centroid of the diamond so that we could establish which solar quadrant the diamond was located in, its angular distance from the E--W direction and the width of the diamond. The thickness of the diamond feature was measured by extending the line from the solar disk centre through the diamond centroid and determining the two points of intersection along the diamond contour -- the distance between those intersecting points provided the diamond thickness. 

This location and angle information were used to rotate the image to the standard format, where the diamond was lined up with the solar E--W direction. Whether to rotate clock-wise or anti-clockwise was determined based on which quadrant the diamond was located in (as discussed previously in Section ~\pref{rotate}). Finally, the DR-phase images were ordered based on the location of the diamond along the ring and its apparent width in the standardised image (as described below, in Section \ref{order}). 

The key elements of the DR image processing steps are illustrated in Figure~\pref{diamond}. The figure on the left shows the original, raw DR-phase image with no processing. In the middle figure the solar disk is identified and fit with a (purple) circle, the diamond was detected and contoured (pink) and a (green) line was drawn connecting the solar disk centre and the diamond centroid to help identify the inclination angle and the diamond width. The figure on the right shows the final result, where the solar disk is rotated to the standard position.

\subsection{Temporal Ordering of the Images}
\label{order}
The images used in this sample have no time-stamp, or the time-stamp was not reliable. Ideally, precise time-stamps would make the ordering of an image sequence elementary, and we hope that by 2017 there will be an element of image time-stamp standardisation in mobile (already in place) and DSLR cameras as the GPS incorporation increases in these devices. However, presently we have to assume that such precise information is, and will not be, available. Hence, we have devised an alternative approach to place the image sequence in a reasonable (if not absolute) temporal order, using the solar crescent and diamond E--W location and widths as a proxy for time during the eclipse. Here we discuss the solar crescent and diamond phase ordering together as the process was very similar in both cases.

\begin{figure}[!t]
\centerline{
\includegraphics[scale=0.48, trim= 0 230 0 230 , clip]{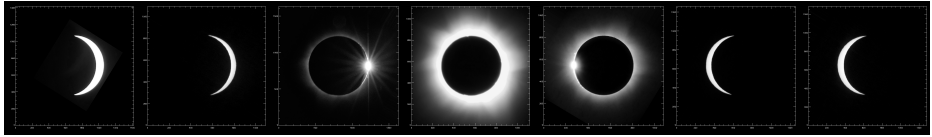}}
\caption{An illustration of the aligned, and time-ordered eclipse observations. (See the link to the animation of this figure in Section \ref{order}.).}
\label{eclipse}
\end{figure}

The first step was to differentiate pre-, and post-totality observations. A diamond or a solar crescent observed on the west side of the solar disk is pre-totality, and the size of the diamond/crescent reduces gradually until totality is reached. The reverse happens on the east side of the solar disk -- post-totality. Hence, the images were separated into two groups -- the west and east limb images. The measured thickness of the diamond/crescent then allowed us to establish the temporal order of the images: descending crescent thickness on the West, followed by descending diamond thickness on the West; totality; and then growing diamond thickness on the East, followed by growing crescent thickness on the East. A demonstration of this ordering process can also be seen in Figure~\ref{eclipse}.

At present, the full-disk or totality images are not possible to order as we have no information on when they were taken (pre-, or post-totality). Nevertheless, we added the full-Sun images at the very beginning and at the very end of the sequence (before and after the start of the eclipse) and inserted our list of total eclipse images between the pre-, and post-totality DR-phase images. This way we created a seamless transition between the different phases of the solar eclipse (see, Figure~\ref{eclipse} and the accompanying animation).

During this project we worked with eclipse photos originating from many different cameras. Some had time-stamps, some did not. For this reason we cannot rely on the availability and accuracy of time-stamps, instead, we used eclipse phase and orientation information to order the images. During the next total solar eclipse (2017) we will obtain an even larger database with thousands of eclipse images including professional, science-quality eclipse photos. We plan to process the public and the professional images separately. Photos taken by the public will be sequenced as described above. However, when processing the professional eclipse photos, we will be able to rely on accurate time-stamps and orientation, which will allow us to create a science-quality, high-resolution eclipse movie, which will benefit the solar physics community as well as the public.

\subsection{Reducing Intensity Flickering}
\label{order}

Once the images were standardised and time-ordered, the proto-movie was created (see \url{http://download.hao.ucar.edu/pub/mscott/SEISS/SEISS\_Movie1.mov}). The flickering seen during the DR and the totality phase is caused by the various quality images produced by cameras of different sensitivity. The goal of our project is to engage the public and use the images submitted as much as possible. Hence, we aim to include a wide variety and quality of images. Nevertheless, we have also tested ways to reduce the intensity flickering. This was done by taking the median intensity of an annulus extending from 1--1.5 solar radius, and using this value to normalise the images to a similar intensity value. This reduced the intensity of the flickering, but meant that the noise in the lower quality images increased making the quality difference more obvious. We solved this issue by removing the lower quality images altogether. As a result the movie became slightly shorter, but the flickering was considerably reduced (see the animation: \url{http://download.hao.ucar.edu/pub/mscott/SEISS/SEISS\_Movie2.mov}). 

Using larger databases in the future, we will be able adjust our quality requirements to use as many good images as possible and still keep the flickering at an acceptable level.

 \section{Results and Conclusions} 
In this article we presented the SEISS tool developed for the standardisation and ordering of multi-source solar eclipse images supplied by the public and professionals. Our goal was to create a robust, and fast tool that is capable of identifying the solar disk at different phases of the eclipse. The images were standardised in resolution, size, orientation and arranged in temporal sequence in order to create a ``megamovie'': a movie of the solar eclipse put together from observations supplied from many sources. 


SEISS is a tool that automatically stitches a heterogeneous set of eclipse images together in a systematic fashion. Currently written in IDL, the code will be ported to a flexible programming language ({\it e.g.} Python) for rapid implementation on web server and portable devices. In this way the EMP citizen scientists will get instant feedback on their movie while their images are uploaded, co-registered and sequenced into the master megamovie. SEISS can also analyse other transit and eclipse events.

With the work presented here, we prepare for the ``Great American Eclipses'' (2017 and 2024) -- events that the whole world will be talking about. They create unprecedented opportunities to educate the public about the Sun, how it impacts the Earth, and also to have them contribute to a very broad scientific experiment that could set a world record for participation and produce the highest time-resolution movie of the Sun's corona ever produced.

\section{Acknowledgements}
\begin{acks}
We would like to thank everyone who kindly shared their eclipse observations with us, making this work possible. This work was supported by the NSF RAPID grant AGS-1247226. We also thank Claire Raftery, Lucia Kleint, and the Eclipse Megamovie Project team for their contribution to the project. Special thanks to Doug Biesecker for helpful discussions.
\end{acks}

%
%
\bibliographystyle{spr-mp-sola} 
\bibliography{eclipse1}  
%
%
%
%

\end{article} 
\end{document}